\pdfoutput=1

\documentclass[11pt]{article}

\usepackage{ACL2023}
\usepackage{booktabs}
\usepackage{times}
\usepackage{latexsym}
\usepackage{xcolor}
\usepackage{graphicx}
\usepackage{listings}
\usepackage{forest}
\usepackage{tikz}
\usepackage{varwidth}
\usepackage{multirow, makecell}
\usepackage[many]{tcolorbox}
\usepackage{caption}
\usepackage{subcaption}
\usepackage{graphbox} %

\definecolor{main}{HTML}{4472C4}    %
\definecolor{sub}{HTML}{EBF4FF}     %

\newtcolorbox{boxA}{
    enhanced, breakable,
    boxrule = 0pt,
    colback = sub,
    borderline west = {2pt}{0pt}{main}, 
    borderline east = {2pt}{0pt}{main}, 
}

\definecolor{mycolor}{rgb}{0.1, 0.6, 0.1}

\newcommand\clustercov[1]{\textsc{ClusTREC-COVID}}
\newcommand\scitoc[1]{\textsc{SciTOC}}
\usepackage[T1]{fontenc}

\usepackage[utf8]{inputenc}
\usepackage{float}
\usepackage{placeins}

\usepackage{microtype}

\usepackage{inconsolata}

\title{Knowledge Navigator: LLM-guided Browsing Framework \\for Exploratory Search in Scientific Literature}

\author{\makecell{Uri Katz$^{1}$~~~~~ Mosh Levy$^{1}$ ~~~~~ Yoav Goldberg$^{1,2}$}\\ 
$^{1}$Bar-Ilan University\hspace{5mm}
$^{2}$Allen Institute for AI \hspace{5mm} \\ 
\texttt{\small\makecell{\{urikacid,moshe0110,yoav.goldberg\}@gmail.com}} 
}

\begin{document}

\maketitle
\begin{abstract}
    The exponential growth of scientific literature necessitates advanced tools for effective knowledge exploration. We present Knowledge Navigator, a system designed to enhance exploratory search abilities by organizing and structuring the retrieved documents from broad topical queries into a navigable, two-level hierarchy of named and descriptive scientific topics and subtopics. 
    This structured organization provides an overall view of the research themes in a domain, while also enabling iterative search and deeper knowledge discovery within specific subtopics by allowing users to refine their focus and retrieve additional relevant documents. Knowledge Navigator combines LLM capabilities with cluster-based methods to enable an effective browsing method. We demonstrate our approach's effectiveness through automatic and manual evaluations on two novel benchmarks, \clustercov{} and \scitoc{}. Our code, prompts, and benchmarks are made publicly available.
\end{abstract}

\section{Introduction}
Traditional search engines, while adept at retrieving relevant documents for specific queries, are sub-optimal when dealing with broad, topical queries. 
Such queries typically return lengthy ranked lists of potentially relevant papers, which, while comprehensive, overwhelms researchers with an information overload, obscuring the underlying structure of the topic and hindering the discovery of relevant subtopics and novel connections. Simply put, researchers are presented with an extensive inventory of documents without a clear map to guide their exploration, while they are interested in understanding broader topical trends.

The limitations of ranked list search results have driven a longstanding interest in methods for grouping and categorizing retrieved documents \citep{kaki2005findex,hearst2006clustering}. Over the years, a vast amount of research was devoted to designing different methods to support the paradigm of grouping and categorizing retrieved documents to organize them into a meaningful structure of knowledge, in many cases taking the form of hierarchical, cluster-based navigation based on automatically induced topical clusters (See \S\ref{sec:exploring}).
However, these browsing methods ultimately did not achieve widespread use and were not adopted by modern search engines, largely due to the insufficient quality of the automatically derived structures for practical application.

\begin{figure}[t!]%
  \centering
  \includegraphics[scale=0.47]{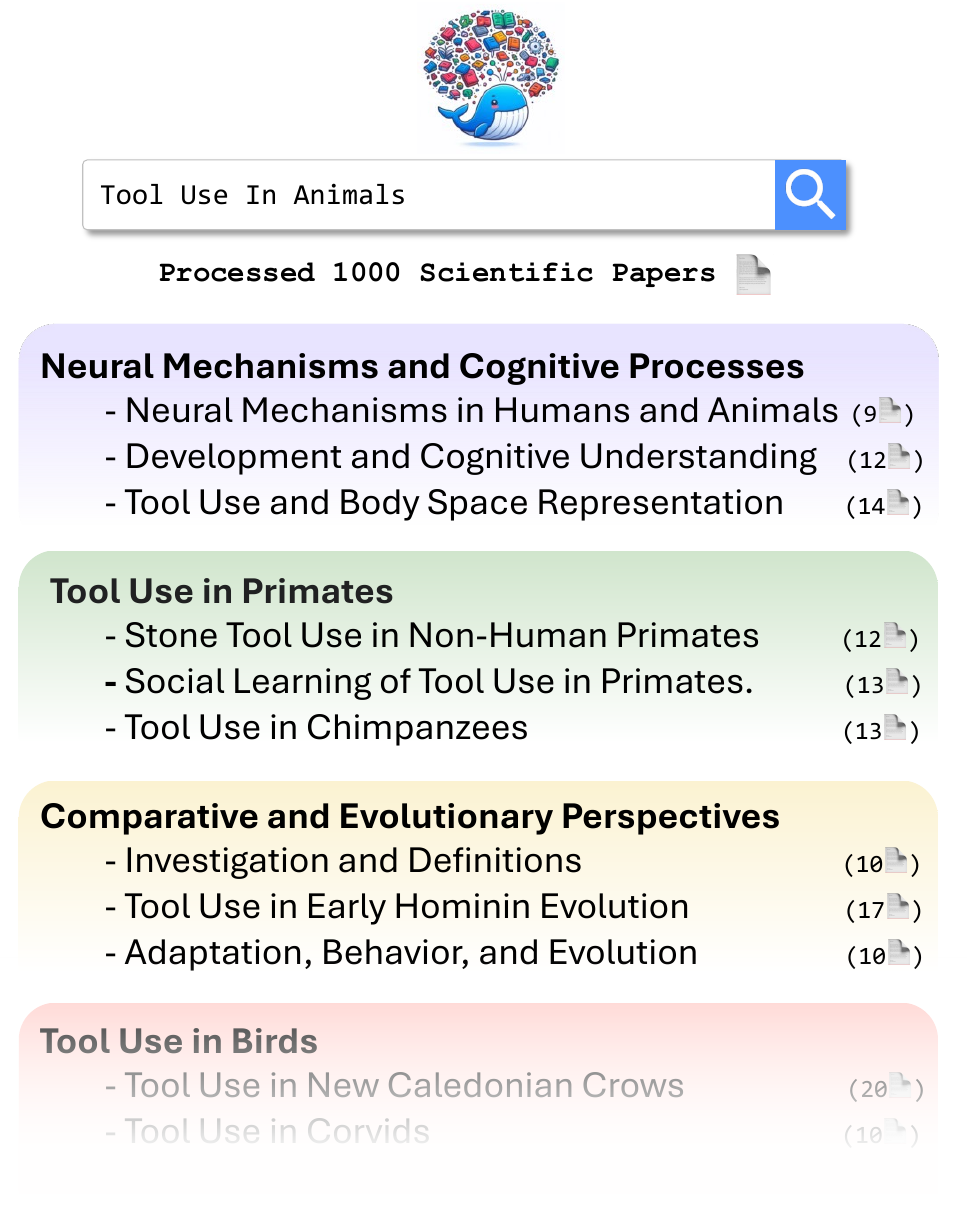}
  \caption{Hierarchical knowledge map generated by Knowledge Navigator, illustrating the primary themes and subtopics identified within a corpus of scientific literature retrieved for the query "Tool Use in Animals." This map demonstrates the system's ability to organize and structure knowledge on a broad topic.}

  \label{fig:knowledge_map}
\end{figure}

In this work, we demonstrate that importing the cluster-based navigation paradigm to the era of large language models (LLMs), combined with modern NLP and IR methods, can overcome many of the obstacles faced by previous approaches. Our research shows that the components required to support this paradigm perform well as stand-alone tasks, and the entire framework functions effectively end-to-end.

\textbf{We propose Knowledge Navigator}\footnote{Knowledge Navigator code, evaluation datasets, and the Streamlit app can be found in \url{https://knowledge-navigators.github.io}}, an LLM-based framework that transforms a large corpus of retrieved scientific literature into multi-level, organized themes of subtopics. 
Given a broad query, Knowledge Navigator generates a list of high-quality subtopics, each accompanied by a readable and interpretable summary grounded in the documents within the corpus. Each subtopic of interest can be expanded through ad-hoc secondary retrieval of fine-grained documents within that specific area. See Figure \ref{fig:knowledge_map} for a graphical illustration of the system's real-world outputs.
These subtopics represent meaningful research clusters, enabling searchers to identify areas of interest, uncover novel connections, and explore specific domains within the broader topic.

By shifting the focus from individual documents to organized subtopic clusters, Knowledge Navigator offers a potential solution to the challenges inherent in traditional ranked list presentations for scientific literature search.

 \textbf{We evaluate} the Knowledge Navigator framework components on various LLMs and representation models, demonstrating its viability with both proprietary models (\textit{GPT-4o}) and open-source models (\textit{Mixtral-8x7B}).
To evaluate the system's components and overall performance, we use (a) \clustercov{}, a modified novel form of the TREC-COVID benchmark \citep{10.1145/3451964.3451965}, which we adapted for subtopic clustering, cluster-based aspect generation, and query generation tasks; and (b) \scitoc{}, a new dataset of scientific review table of contents, constructed from "Annual Reviews" open-access journals in a variety of scientific fields.  We publish those datasets for future work in  NLP research. 

Using these benchmarks in both automatic and domain expert evaluation, we demonstrate that Knowledge Navigator performs efficiently in each of its component tasks, as well as in its overall function of organizing and outlining scientific knowledge. To our knowledge, this work is the first to showcase the feasibility of modern LLMs for supporting cluster-based navigation paradigms and aims to contribute to both the research and development of modern browsing systems.

\section{Exploratory Information Seeking through Knowledge Navigation}
\label{sec:exploring}
When scientists explore a new topic and review the literature, their information-seeking behavior is an instance of “exploratory search” \citep{meho2003modeling,Soufan_Ruthven_Azzopardi_2022}.
That is when a searcher does not have a particular document or topic in mind but rather is interested in finding out the different aspects reflected in the \emph{document collection}, to both gain an understanding of the overall structure of the domain, as well as to look for subtopics that may interest them and/or fit their expertise and interests \citep{Marchionini_2006,white2009exploratory}.

Currently, this process is not well supported by modern search systems, although it has been found to be a common search behavior among scientists \citep{Nedumov_Kuznetsov_2019,tahri-etal-2023-transitioning}. A major obstacle is the inability of searchers to effectively consume hundreds to thousands of documents and distill topics from them. Rather, they browse tens of document titles at a time, revise their mental model of the domain based on this subset, maybe take notes,  adapt their query to reflect their new mental image and interests, and navigate into a specific aspect or subtopic of interest.

Knowledge Navigator enhances this exploration process by consuming hundreds of results, finding common themes, and organizing them into a two-level hierarchy of subtopics, allowing users to systematically explore different facets of a broad topic, thereby addressing the complex and multifaceted nature of their information needs. 
By transforming search results into organized subtopic clusters, Knowledge Navigator supports a holistic way of absorbing new information, helping them to identify areas of interest and discover novel connections even when their queries are initially vague or evolving.

Cluster-based browsing has been extensively studied in the past \citep{Cutting1992ScatterGatherAC,hearst1999use,Zamir1999GrouperAD,Osinski2005ACA}, but despite this, these methods were not widely adopted. Approaches like Scatter/Gather \citep{Cutting1992ScatterGatherAC}, which aimed to organize documents into coherent clusters for easier navigation, failed to produce good representations of documents in practice, leading to clusters that did not accurately differentiate between subtopics \citep{hearst1999use}. Additionally, these methods struggled to generate clusters that were easily interpretable by users due to reliance on keyword extraction techniques that were difficult for searchers to understand \citep{Zhang_Broussard_Ke_Gong_2014}. This was largely due to the immature state of Information Retrieval (IR) and Natural Language Processing (NLP) methodologies at the time.

The introduction of instruction-tuned LLMs, with their advanced world knowledge and text understanding abilities \citep{ouyang2022training,achiam2023gpt}, has recently led to renewed research utilizing LLMs for information organization across various tasks \citep{pham2023topicgpt,viswanathan2023large,Zhang2023ClusterLLMLL}. We demonstrate that LLMs can be effectively used in assisting scientific literature consumption, by organizing document collections that result from a query into a digestible "table-of-content" of the topic, where each sub-topic is grounded in concrete research works. This builds on the parametric knowledge of the LLM about scientific concepts, their categorization, the relations between them and many other facets of scientific knowledge gained in training, but relying on this knowledge in a fully grounded way, using it solely for the purpose of cataloging and organizing a given set of human authored documents, which result from a provided query.

\section{Knowledge Navigator}
\begin{figure*}[htbp]

  \centering
  \includegraphics[scale=0.43]{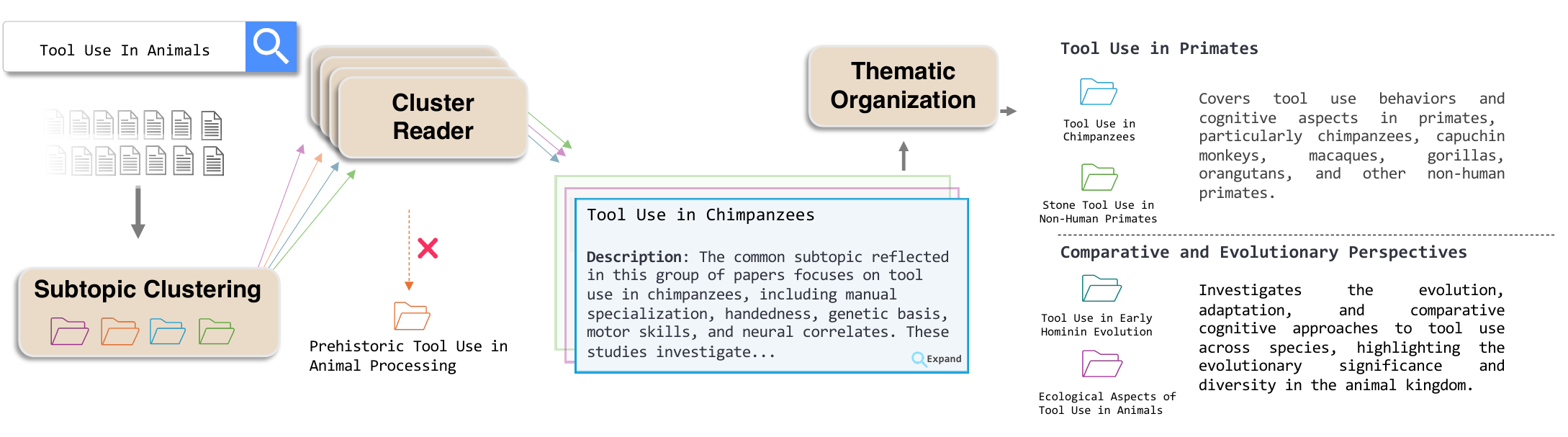}
  \caption{Knowledge Navigator Workflow: Starting with a query to a scientific literature retriever (e.g., Google Scholar), retrieved documents are embedded and clustered. The Cluster Reader then generates descriptive titles and descriptions for each cluster and filters for relevance. Finally, the Thematic Organization module groups the subtopics into a structured outline}
\label{fig:arch}
\end{figure*}
Knowledge Navigator system takes a corpus of retrieved scientific documents for a given query and outputs an organized two-level thematic structure of subtopics spanning that topical query. The system's functionality is supported by the following conceptual steps: corpus construction, embedding and clustering of documents, describing and naming clusters, filtering irrelevant clusters, grouping the clusters into a thematic hierarchy, and subtopic query generation. This is implemented in a five-component architecture that largely follows the conceptual steps.

This architecture enables LLMs to generate grounded outputs based on a large number of source documents, a crucial requirement for organizing and structuring large corpora.

\paragraph{System Design}
LLMs are incredibly effective at consuming information but are expensive to run and bottlenecked by the amount of information they can effectively process in their prompts. Thus, we design the system around these constraints, attempting to minimize the number of LLM calls and aiming to make the most effective use of each one. Each step is designed to reduce the information size and transform it into a suitable form to be fed as input to the next LLM step.
The system is thus designed to work bottom-up, progressively abstracting information at each stage. 

Starting from hundreds of search-query results, represented as titles, abstracts and snippets, we employ a relatively cheap operation of contextual embeddings followed by clustering, to organize them into smaller---but cohesive---groups. Each group is then fed, as a whole but separately from other groups, into the Cluster Reader, an LLM-based component that is in charge of analyzing it, describing its common themes, naming it, and scoring its relevance to the query. Then, the names and descriptions of all the relevant clusters are fed (together with the initial query) into a subsequent LLM-based component, which organizes them into thematic groups, and names these groups. 
The result is a corpus-level hierarchical organization of the search results, which can serve as a map to the scientific topic, and whose construction relied on the entire corpus: the clustering is a global operation; the naming is separate per cluster but takes the query into account and considers many documents in assigning the description and name; and the second-stage thematic organization again provides a global view based on all the clusters who passed the relevance filter. 

The searcher can then browse the generated topical outline, identify a sub-topic of interest, and initiate the Subtopic Expander to automatically generate a query to retrieve additional documents on the fine-grained sub-topic.

We now describe each component.

\subsection{Topical Corpus construction}
\label{subsec:topic_construction}
The initial step starts with a search query reflecting a relatively broad scientific topic $T$ (e.g. "Tool use in animals"), and results in a topical corpus \(C\) comprising the top \(K\) documents ranked by a search engine for this query.\footnote{In our research, we utilize Google Scholar via \href{https://serpapi.com/}{SerpAPI service}, though the method is not confined to any specific search engine or retrieval technique as long as they are capable of retrieving a pool of relevant documents.}
We select a large \(K\) (up to 1000) to ensure a diverse set of research papers that represent the full spectrum of the topic. 

\subsection{Subtopic Clustering}
\label{subsec:subtopic_cluster}

Next, we aim to divide the corpus $C$ into subtopic groups, each reflecting a sub-topic $t_i$ of the broad topic $T$. This is done via clustering of contextual embedding vectors. We represent each retrieved document as a single embedding vector derived from the paper's title and its snippet and abstract (section \ref{eval:cluster} compares various embedders).

For clustering, we use Gaussian Mixture Model (GMM), a probabilistic soft-clustering algorithm that can assign each sample to one or more clusters.

The optimal number of clusters is determined using the Silhouette score \citep{10.1016/0377-0427(87)90125-7}, balancing cohesiveness and separation without penalizing the complexity of the GMM. 
Given the high-dimensional nature of modern text embeddings, and relatively low sample count, clustering methods face challenges in accurately assessing sample proximity, often resulting in poor quality clusters \citep{aggarwal2001surprising}. 
To address this, prior to clustering, we reduce the dimensionality of the vectors using UMAP (Uniform Manifold Approximation and Projection) \citep{mcinnes2018umap}. %

\subsection{Cluster Reader}
\label{subsec:cluster_reader}
The Cluster Reader is an LLM-based component that operates independently on each subtopic cluster, receiving as input the titles and abstracts of all documents within that specific cluster, along with the initial topical query. This component serves two functions, --naming and filtering--, which are achieved in a single prompt (See \ref{app:prompt_cluster_reader} for the exact prompt). Its output is a subset of the clusters, each with an associated name and description.

\paragraph{Describing and Naming Subtopics}
The Cluster Reader first reads the initial query, paper titles, and abstracts within each cluster to identify and articulate the specific subtopic they address. It generates a detailed description that encapsulates the thematic essence of the cluster in relation to the broader topic (T). Based on this description, it then generates a meaningful title for the subtopic. The output of this process is a detailed description of the subtopic, along with a subtopic title. The performance of the subtopic naming function is evaluated by a domain expert in \S\ref{subsec:eval_naming}.

\paragraph{Subtopic Filtering}
In the same LLM call, after the Cluster Reader generates a description and title for each subtopic cluster, it then scores the subtopic's relevance to the original topic T on a scale of 1 to 5. Based on this score, it determines whether to filter out the subtopic cluster. This filtering process eliminates clusters deemed explicitly unrelated to topic T, addressing the noise often present in large retrieved document collections on broad topics. The performance of this filtering function is evaluated by a domain expert in \S\ref{eval:filter}.

We chose to implement the three steps (naming, describing, and filtering) of the Cluster Reader as a single LLM call to induce a scratchpad reasoning process \citep{nye2021show}, where each step builds upon the previous stages. This also motivated the ordering of the generated content, from a detailed description to a concise name, to relevance scoring, and finally to relevance judgment.

\subsection{Thematic Organization}
\label{subsec:outline_creation}

The set of subtopics resulting from the cluster reader are diverse and fine-grained. 
The Thematic Organizer component takes all of the subtopic names and descriptions as inputs and groups them into meaningful thematic groups. 
For example, \textit{"Dinosaur Thermoregulation and Metabolism"} and \textit{"Dinosaur Musculature and Biomechanics"} would be grouped under \textit{"Physiology and Functional Morphology"}, while \textit{"Evolutionary Transition from Dinosaurs to Birds"} and \textit{"Origins and Ascent of Dinosaurs"} grouped under \textit{"Evolution and Phylogeny"}. Such an organization greatly helps in browsing the list of results.

The cluster names and descriptions are sufficiently short for their entire set to fit in a single prompt, which is how the thematic organizer is implemented. 
The prompt contains the full set of topics and an instruction to organize them into higher level groups.

From a technical standpoint, we found it essential to associate each input topic with an explicit numeric ID and task the LLM with listing the IDs for each of its generated high-level themes, rather than to replicate the cluster names in its output. Using the IDs greatly reduced hallucinations, ensured consistent output, and increased coverage of the input topics. See \ref{app:prompt_thematic_org} for the exact prompt.

\subsection{Subtopic Expander}
The Subtopic Expander is an LLM-based component designed to enable searchers to automatically retrieve additional scientific documents relevant to fine-grained subtopics, allowing for deeper exploration of a specific subtopic without the need for manual query curation. Given the content of a subtopic cluster (i.e., its title, description, and assigned papers), the Subtopic Expander generates a list of terms directly related to the subtopic and specifically extracted from the scientific terminology present in the cluster's papers. These extracted terms are then concatenated into a single query, which is used to retrieve additional results that will populate the subtopic cluster. The expanded subtopic cluster can subsequently serve as an initial topical corpus for a secondary organization into a two-level thematic structure of subtopics. We conduct an isolated evaluation of the Subtopic Expander on \clustercov{} in \ref{subsec:clustercov_exp}, and an end-to-end system evaluation on \scitoc{} in \ref{subtopic_expander_scitoc}. See \ref{app:prompt_subtopic_expander} for the exact prompt.

\section{Structured Scientific Literature Benchmarks}

An ideal dataset for evaluating the system end-to-end would necessitate exhaustive annotation of thousands of scientific documents, classifying each for relevancy, subtopics, and thematic groups. Unfortunately, no existing dataset supports this task comprehensively. To this end, we introduce two novel benchmarks, \clustercov{} and \scitoc{}, that facilitate a robust evaluation of Knowledge Navigator, both on an individual component level and end-to-end.\footnote{Benchmarks are publicly available in our GitHub project.}

\paragraph{\clustercov{}}
To assess the construction and naming of subtopic clusters, we modified TREC-COVID, which was initially designed as an information retrieval (IR) benchmark for evaluating search performance on scientific literature related to COVID-19  \citep{10.1145/3451964.3451965}. It includes 50 expert-curated queries, each representing a subtopic of COVID-19 research across multiple fields, with each query serving as a concise subtopic title, such as 'coronavirus heart impacts'.
For each query, medical experts judged hundreds of documents, annotating each for relevance to the topic, resulting in an average of 300 highly relevant documents per query.
These characteristics—expert curation, diverse subtopics, and detailed relevance annotations—make the TREC-COVID benchmark particularly suitable for evaluating the Knowledge Navigator system’s components.
We transformed the TREC-COVID benchmark into a clustering benchmark by forming clusters from documents annotated as "highly relevant" to specific topics. 
We randomly sampled up to 50 documents from each topic, ensuring documents appear in only one cluster. 
This resulted in a dataset with 2,284 documents assigned to 50 subtopic clusters labeled with expert-curated titles.

\paragraph{\scitoc{}}
To assess the system's ability to handle complex scientific topics and produce well-organized outputs, we constructed a novel benchmark of 50 tables of contents (TOC) of scientific reviews sourced from 15 diverse peer-reviewed journals published by Annual Reviews\footnote{\url{annualreviews.org}}. These reviews span a wide array of scientific fields, including biology, medicine, food industry, environmental studies, and more. Reviews were selected based on predefined criteria: explicit and scientifically described tables of contents and subtopics, excluding metaphorical or abstract language.  For instance, the review paper \textit{The Effects of Psychedelics on Neuronal Physiology} \citep{hatzipantelis2024effects} included headers such as "Effects of Psychedelics on Gene and Protein Expression" and "Effects of Psychedelics on Neuronal Survival and Neurogenesis"---which we keep---as well as structural headers like "Introduction," "Background," and "Discussion,"---which we discard. See an example from \scitoc{} in  \ref{scitoc_example}. 

For each review paper, we transformed the main title into a query focused on scientific terms to retrieve documents relevant to the review topic. Queries included only keywords appearing in the review title. For instance, the aforementioned title was modified to the Boolean query \texttt{"Psychedelic" AND "Neuronal Physiology"} to enable accurate retrieval of documents related to the main topic of the review.

\section{Experiment}
\label{sec:comp_eval}
The Knowledge Navigator system comprises several components that have not been extensively explored, particularly within the domain of scientific literature. %
We present an evaluation of these components, both individually and as part of the integrated system, specifically focusing on the organization of topical scientific corpora.

\subsection{\clustercov{} Experiments}
\label{subsec:clustercov_exp}

\noindent\textbf{%
Does clustering effectively discover subtopics in an already topical corpora? }
\label{eval:cluster}
Clustering algorithms aim to group similar instances, but their success in accurately clustering scientific documents into coherent subtopics is uncertain due to the complex nature of scientific concepts, which may not align with the algorithms' similarity measures.
Our goal is to assess whether the clustering component effectively organizes \clustercov{} documents into subtopic clusters that align with human annotations..

For the evaluation of the subtopic clustering component, we experiment with GMM clustering using various text representation methods. 
Following literature on document clustering in information retrieval \citep{Yuan_Zobel_Lin_2022}, we report relevance-based measures. 
We use "Clusters per topic by relevance" ($R_{c}@p$) metric which indicates the number of clusters needed to observe p\% of relevant documents, while "coverage per topic by relevance" ($R_{v}@p$) shows the \% of minimum number of documents needed to cover p\% of relevant documents.

Table \ref{tab:clustering} shows that all models reconstructed topic groups with a fair degree of correlation between the gold mapping and clustered map. 
The best results were achieved by \textit{text-embedding-3-large} and \textit{SFR-mistral}. The $R_{c}@80$ metric indicates that, on average, a searcher needs to evaluate 2.3-3.18 clusters out of 50 to cover 80\% of the documents in a topic, compared to an average of 20.4 clusters in a random cluster assignment, and 5-7\% of the entire collection (See $R_{v}@80$) compared to 41\% of the corpus documents. Interestingly, the \textit{SPECTER2} model, despite being trained for scientific documentation, lagged behind general text embedding models. For the rest of the evaluations, we used the \textit{text-embedding-3-large} model for ease of use via an API, but the results indicate that an open-source version would achieve similar performance.
\begin{table*}[t]
\centering
\small %
\begin{tabular}{lp{3.1cm}p{2.5cm}p{2.5cm}}
\toprule
 & \textbf{Adjusted Rand Index $\uparrow$} & \textbf{NMI$\uparrow$} & \textbf{$R_{c}@80$ / $R_{v}@80$ $\downarrow$} \\
\midrule
\textbf{text-embedding-3-large} & \textbf{0.516} & \textbf{0.732}  & 2.4 / 5.1\% \\
\textbf{text-embedding-3-small} & 0.496 & 0.719  & 2.5 / 5.7\% \\
\textbf{SFR-Mistral 7b} & 0.513 & 0.736  & \textbf{2.3 /5.1\%} \\
\textbf{SPECTER2} & 0.435 & 0.674  & 3.18 / 7\% \\
\textbf{Random} & 0.00 & 0.153 &  20.4 / 41\% \\ 
\bottomrule
\end{tabular}
\caption{Comparison of different vector representations for clustering scientific documents to subtopics in the \clustercov{} benchmark}
\label{tab:clustering}
\end{table*}

\noindent\textbf{Can LLMs accurately identify and name underlying subtopic in a document cluster? }
\label{subsec:eval_naming}
We evaluate the cluster reader's ability to replicate the cluster names assigned to \clustercov{} clusters by expert annotators.
The Cluster Reader received the titles and abstracts of all the papers in a cluster as input and was tasked with generating a representative title for that cluster. Each generated title was then judged against the original subtopic query for a match, with the evaluation conducted by annotators with academic backgrounds in biomedical research.

Table \ref{tab:aspect_generation} shows that the Cluster Reader (using \textit{GPT-4o} or \textit{Mixtral-8x7B}) successfully generated subtopics that matched 88\% of the subtopics in \clustercov{}.  
Most unmatched subtopics were closely related to the benchmark's subtopics but not identical (e.g., "COVID-19 in African-Americans" (expert) vs. "Racial and Ethnic Disparities in COVID-19 Outcomes" (cluster-reader)). 
When papers were assigned to random clusters, the generated subtopic titles were mostly broad and general descriptions of COVID-19 research, resulting in very low coverage (6\%). None of the subtopics were filtered by the Cluster Reader, as desired.
        
\begin{table}[h]
    \centering
    \small
    \begin{tabular}{lc}
        \toprule
        \textbf{Experiment} & \textbf{\% Subtopic Match} \\
        \midrule
        Random clusters + GPT-4o &$6\%\pm{3}$ \\
        GPT-4o & $88\%\pm{4}$ \\
        Mixtral-8x7B & $88\%\pm{4}$ \\ 
        \bottomrule
    \end{tabular}
    \caption{Comparison of Generated Subtopic Titles to Ground Truth in \clustercov{}}
    \label{tab:aspect_generation}
\end{table}

\noindent\textbf{Can LLMs generate effective queries from clusters of scientific documents? } 
To evaluate the effectiveness of the Subtopic Expander in generating queries from scientific document clusters, we created 50 subtopic clusters, each containing 20 randomly sampled relevant papers from one of the 50 topics in \clustercov{}. The Subtopic Expander then used the subtopic title, description, and associated papers to produce a specialized query. We assessed retrieval performance using the BM25 retriever against the entire TREC-COVID corpus (192K documents). We compared the Subtopic Expander’s performance against two baselines: (1) generating a new query based solely on the subtopic title without considering the papers in the cluster, and (2) using the original, unmodified TREC-COVID topic query. Remarkably, the queries generated by the Subtopic Expander significantly outperformed both baselines (see table \ref{tab:query_summary}), with improvements of up to 7.4\% in precision@K and 14.2\% in recall@K (see Appendix for details on recall@K) compared to the original TREC topic queries. These results demonstrate that this method enables searchers to achieve superior retrieval capabilities in fine-grained scientific subtopics without requiring prior knowledge of specific terms or jargon embedded in the subtopic cluster’s papers.

\subsection{\scitoc{} Experiments}
We now present our evaluation based on the Scientific Reviews Table-of-Contents benchmark. Unless otherwise specified, the evaluations are based on a complete human annotation of 20 reviews, annotated by a hired academic researcher with a PhD in Biology. The annotator evaluated each generated subtopic in these 20 scientific review papers, resulting in a total of 1,471 relevant subtopics and 261 filtered subtopics. Each subtopic was assessed through multiple questions, which will be detailed in the following subsections. We performed an inter-annotator agreement assessment on a subset of reviews to evaluate the reliability of the annotation process. The results indicated a high level of agreement for the tasks evaluated. Further details are provided in \ref{app:agreement}.

\paragraph{Can LLMs effectively filter irrelevant subtopic clusters?}
\label{eval:filter}
To assess the Cluster Reader's ability to filter non-relevant subtopics, the expert annotator evaluated the relevance of each generated subtopic title and summary to the original query topic. Relevance was defined as the subtopic having a direct and clear connection to the original topic. The annotation process covered both filtered and non-filtered subtopics to assess filter performance.
The Subtopic Filter flagged 261 subtopic clusters as not relevant to the initial topic. Of these, 87.7\% (229) were confirmed as non-relevant by the annotator. Conversely, only 0.14\% (2 subtopics) of the 1471 non-filtered subtopics were marked as non-relevant. This indicates that the filter has a high precision and a very low false negative rate, resulting in an overall accuracy of 98.8\%. effectively removing irrelevant subtopics while retaining those that are relevant.
\paragraph{Can LLMs organize subtopics into coherent thematic categories?}
\label{eval:outline}
To assess the thematic organization component, we evaluated the Knowledge Navigator's output for each topic in the Scientific Reviews benchmark. For each of the 1,471 generated subtopics, an expert annotator determined whether the subtopic was assigned to a relevant theme. For example, in the topic "Gut Microbiota in Colorectal Cancer," the subtopic \textit{"Role of Probiotics in Colorectal Cancer Prevention and Treatment"} was correctly assigned to the theme "Therapeutic and Preventive Approaches Targeting Gut Microbiota," while the subtopic \textit{"Impact and Modulation of Gut Microbiota in Colorectal Cancer"} was marked as a false assignment. Overall, we found that 
\textbf{95.2\%} of subtopics were assigned correctly to themes within their respective topics.

\begin{table}[t]
\centering
\small
\begin{tabular}{lp{0.0001cm}p{0.4cm}p{0.4cm}p{0.5cm}p{0.5cm}}
\toprule
\textbf{Query type} & \textbf{P} & \textbf{@20} & \textbf{@70} & \textbf{@100} & \textbf{@200} \\
\midrule
Original Query & & 0.43 & 0.36 & 0.33 & 0.27 \\
Subtopic Expander\\
\hspace{3mm}Title & & 0.45 & 0.38 & 0.36 & 0.30 \\
\hspace{3mm}Title + Cluster & & \textbf{0.50} &  \textbf{0.43} & \textbf{0.41} &  \textbf{0.33} \\
\bottomrule
\end{tabular}
\caption{Precision@K on TREC-COVID retrieval using different query generation methods for subtopic expansion}
\label{tab:query_summary}
\end{table}

\subsection{End-to-End Subtopic Coverage Evaluation}
We evaluate the Knowledge Navigator's overall performance to identify meaningful subtopics within diverse scientific domains by comparing its generated output to the table of contents (TOC) of corresponding human-authored review articles. Our assessment encompasses two key aspects: (a) the extent to which the generated subtopic titles match and cover the headers in the reviews TOC (subtopic coverage), and (b) the extent to which the generated subtopics introduce additional topics not present in the human-authored review (novelty of subtopics).

\paragraph{Overall Statistics}
The human-authored reviews have 10.5$\pm{1}$ valid subtopics on average, while the Knowledge Navigator produces an average of 73.5$\pm{3}$ relevant subtopics per topic. 

\paragraph{Automatic Evaluation}
To assess system capabilities across the entire benchmark of 50 review papers for both \textit{GPT-4o}\ and \textit{Mixtral-8x7B} \citep{jiang2024mixtral}, we employed a heading soft recall \citep{franti2023soft} evaluation method, as suggested in \citep{shao2024assisting}, to compare the recall of generated subtopics against human-authored review outlines. This method is suitable due to the comparison of subtopic title lists against an existing review table of contents. 

As a baseline, we prompted the LLM directly to generate subtopic lists for the same topics given to Knowledge Navigator, assessing its ability to generate meaningful scientific subtopics based on its parametric knowledge. It's important to note that the evaluated reviews are publicly available and may have been encountered during the LLM's pretraining.

Table \ref{tab:soft_recall} shows that Knowledge Navigator outperforms the Direct Generation (Direct Gen) setup in both models\footnote{Both pairs show statistically significant differences in the paired t-test with $p < 0.01$}, with a 13\% improvement in coverage for \textit{Mixtral}. This suggests that Knowledge Navigator can compensate for limitations in model scale, enhancing the identification of relevant subtopics. As we see in the next section, the knowledge navigator also produces many novel topics, not covered by either the review or the LLM.

Notably, \textit{Mixtral-8x7B} and \textit{GPT-4o} achieve similar results for Knowledge Navigator (KN in the table), suggesting open-source models can be viable alternatives. 

\paragraph{Domain Expert Evaluation}
We also grounded the automatic evaluation with full human expert evaluation over 20 of the reviews.
To assess Knowledge Navigator's ability to identify meaningful scientific subtopics within a given topic, we compared its generated output to the corresponding review paper's Tables of Contents (TOCs). The review paper's TOC serves as an indicator of the foundational subtopics a reader should encounter. The annotator first identified all relevant headers in the review TOC, excluding headers of general background information unrelated to the topic or subjective headers that do not represent explicit subtopics (see \ref{app:non_valid_headers}). Next, for all valid headers, the annotator matched generated subtopics to TOC headers only if they explicitly addressed the same subtopic. Unmatched yet relevant subtopics were classified as novel.

On average, Knowledge Navigator explicitly covered $71.6\%\pm{3}$ of the review headers and generated $35\pm{4}$ novel subtopics per review (on top of the 10 topics in the review). This result demonstrates the system's ability to identify scientifically meaningful subtopics considered foundational by the experts who authored the reviews. Ultimately, the system aims to organize corpora of scientific literature into a structure that mirrors how an expert would approach the task, while being more exhaustive in its inclusion of relevant subtopics and themes for a comprehensive overview of the broader topic.

\begin{table}[t]
    \centering
    \small
    \begin{tabular}{lc}
        \toprule
         & \textbf{Soft Heading Recall} \\
        \midrule
        GPT-4o Direct Gen & 82.6\% \\
        GPT-4o + KN & \textbf{87.1\%} \\
        \midrule
        Mixtral-8x7B Direct Gen & 75.3\% \\
        Mixtral-8x7B + KN & \textbf{88.3\%} \\ 
        \bottomrule
    \end{tabular}
    \caption{Results for automatic evaluation of subtopic coverage of \scitoc{} tables of contents for Knowledge Navigator (KN) and direct generation by \textit{GPT-4o} and \textit{Mixtral-8x7B}.}
    \label{tab:soft_recall}
\end{table}

\label{subtopic_expander_scitoc}
\subsection{Expanding Subtopics by Retrieving Additional Relevant Papers} We evaluated the Subtopic Expander as part of the entire Knowledge Navigator on the expert-annotated reviews to assess its capabilities in generating queries for the expansion of fine-grained scientific subtopics. We expanded 40 random subtopics, 2 from each of the 20 annotated topic reviews. Using the generated query in a Boolean form where each keyword is concatenated with an "OR" clause, we retrieved 100  documents from the search API. Overall, we collected 4,000 scientific papers for 40 different subtopics.

In order to conduct a relevant judgment evaluation over thousands of retrieved documents, we constructed and validated an LLM-judge capable of assessing a retrieved document’s relevance to the subtopic title. The LLM-judge achieved a high level of agreement with the human expert. See \ref{app:llm_judge} for a detailed description.

\begin{figure}[t]

  \centering
  \includegraphics[scale=0.35]{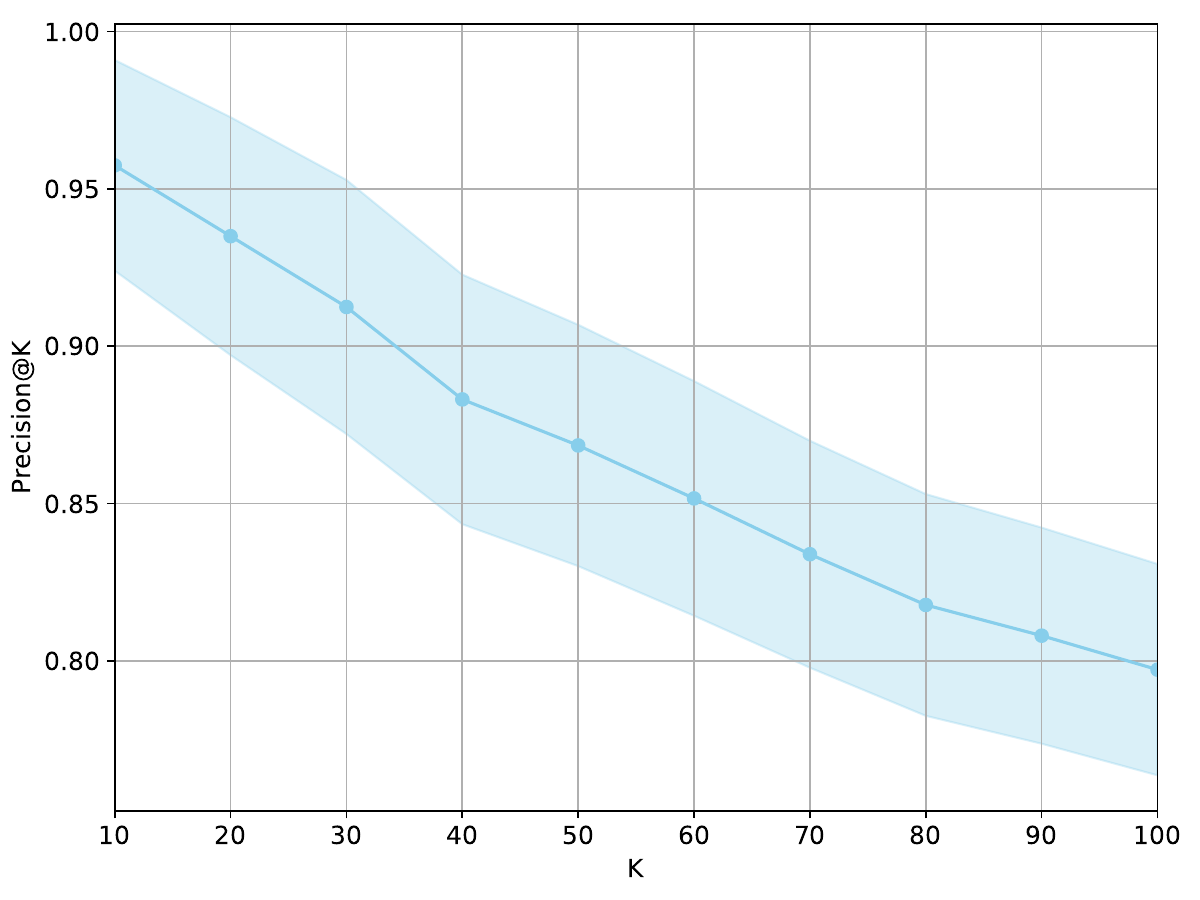}
  \caption{Average Precision@K of the K documents retrieved using a query generated by the Subtopic Expander for the \scitoc{} reviews.}

\label{fig:precision}
\end{figure}

\paragraph{Subtopic Expander Evaluation}  Using the validated LLM judge, we assessed the relevancy of retrieved papers to evaluate Precision@K. The order of the retrieved papers reflects the original sequence from the search API output. As shown in Figure \ref{fig:precision}, the results indicate that the retrieved documents achieve high precision, enabling the accurate retrieval of up to 100 new scientific papers within fine-grained subtopics without requiring the searcher to formulate a new query while maintaining reasonable precision. This demonstrates the system’s potential for scientific exploration.

\section{Related Works}

\paragraph{Cluster-based Browsing} notably the Scatter/Gather paradigm \citep{Cutting1992ScatterGatherAC,pirolli1996scatter}, was proposed to enhance exploratory search \citep{Gong_Ke_Khare_2012} by grouping retrieved documents into clusters and allowing iterative refinement. However, limitations in representing cluster content with extracted keywords hindered its adoption \citep{Zhang_Broussard_Ke_Gong_2014}. Since then, other notable methods have proposed different cluster-based approaches aimed at improving the computational efficiency of clustering algorithms or the selection of representative keywords \citep{Zamir1999GrouperAD,Osinski2005ACA}. Our work draws inspiration from this approach but leverages advancements in LLMs to structure and interpret documents, enhancing interpretability. Additionally, our focus on enabling exploration through multi-level subtopic hierarchies differentiates our system from Scatter/Gather's document retrieval focus.

\paragraph{Information Organization with LLMs}
It was shown that LLMs are capable of clustering items \cite{viswanathan2023large,Zhang2023ClusterLLMLL,wang2023goal} and uncovering latent topics in text collections \cite{pham2023topicgpt}. However, these methodologies do not integrate those capabilities within practical applications, focusing instead on evaluating the capabilities of LLMs in isolation.
In our work, we showed how using LLMs as a component within a framework can transform large corpora of scientific literature into a thematic organization of subtopics. Each subtopic can then be expanded by using an automatically constructed query to retrieve additional relevant documents.

\section{
Discussion
}
We demonstrate how the challenge of navigating the scientific literature when embarking on a new field can be facilitated by an LLM-aided process, which we call Knowledge Navigator.

The Knowledge Navigator operates on a corpus of documents which enables a more holistic understanding and organization of knowledge within the domain.
The effectiveness of our framework demonstrates the potential of the bottom-up approach in other settings where LLMs are tasked with extracting insight from a collection of items.

In addition, we believe that future work could use outputs from frameworks like Knowledge Navigator in prompts for other systems or in planning tasks for agents. 
For example in the retrieval-augmented generation (RAG) settings, where structured data boosts the performance and utility of LLMs in various applications.

\section*{Limitations}
Knowledge Navigator demonstrates promising results in organizing and structuring scientific knowledge, offering a potential solution to the challenges of information overload in exploratory search. However, like any system, it has limitations that can be addressed in future work:
\paragraph{Corpus Quality and Recall.} The system's performance is inherently dependent on the quality of the retrieved corpus. Suboptimal retrieval can still impact the system's output, even with the subtopic filtering mechanism in place. This limitation highlights the importance of further refining the retrieval process to improve recall and ensure the inclusion of all relevant documents.
\paragraph{Document Assignment.} Although Knowledge Navigator utilizes soft clustering to potentially assign documents to multiple subtopics, this approach is not exhaustive due to the limitations of clustering algorithms and the representation space. Exploring alternative assignment strategies could enhance the system's ability to represent complex relationships between documents and subtopics.

\paragraph{User Interface and Experience.} Our work primarily focuses on the technological and system design aspects of information organization.  The development of a user interface (UI) that leverages Knowledge Navigator's capabilities and optimizes the user experience is crucial for its practical application.

\section*{Acknowledgements}
We would like to thank SerpAPI for generously granting us credits for our evaluations.\\
This project has received funding from the European Research Council (ERC) under the European Union's Horizon 2020 research and innovation programme, grant agreement No. 802774 (iEXTRACT).

\bibliography{anthology,custom}
\bibliographystyle{acl_natbib}
\clearpage
\appendix

\onecolumn
\section{Knowledge Navigator interface simulation}
\label{sec:appendixB}
\begin{figure}[H]
  \centering
  \includegraphics[scale=0.5]{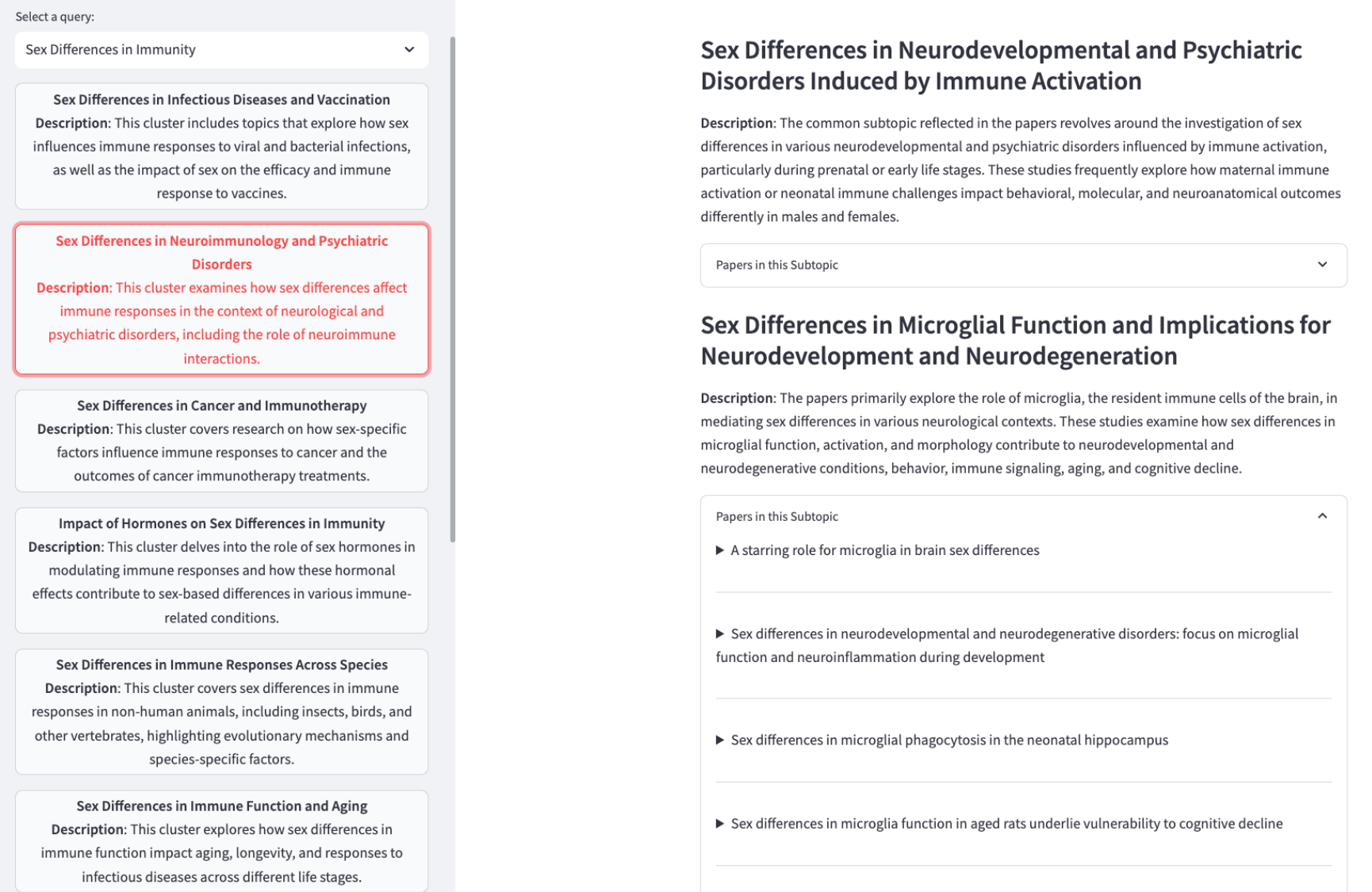}
  \caption{Knowledge Navigator implementation over a Streamlit web application for demonstration}
  \label{fig:streamlit}
\end{figure}
\FloatBarrier
\twocolumn

\onecolumn
\section{Expert Annotation}
\label{sec:appendixC}
\begin{figure}[H]
  \centering
  \includegraphics[scale=0.47]{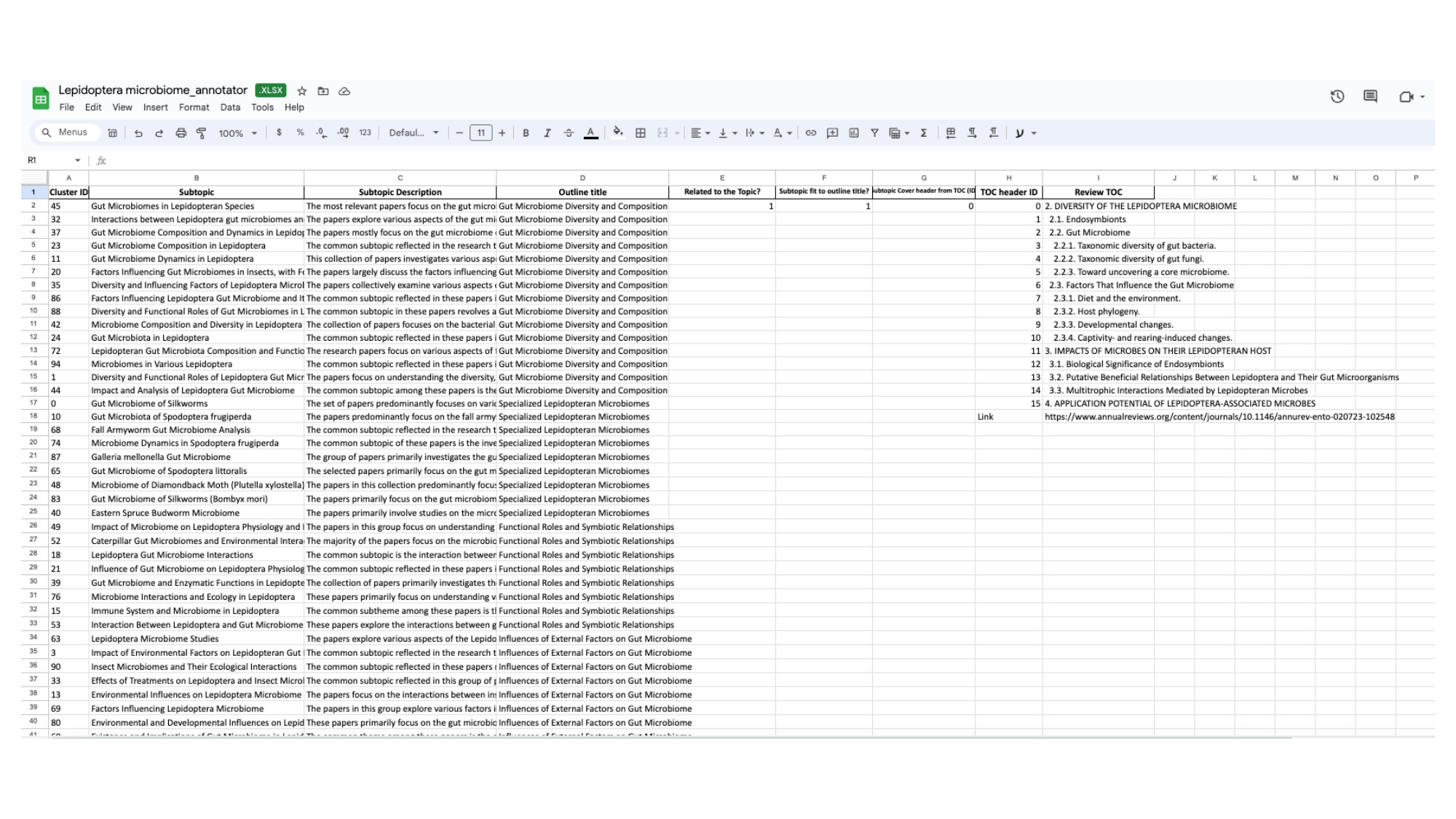}
  \caption{Annotation interface (Google sheet) for the expert annotator. After the training session and overview of the instructions, the annotator evaluated each topic in a separate file.}
  \label{fig:annotation}
\end{figure}
\FloatBarrier
\subsection{Annotator agreement}
\label{app:agreement}
To assess the reliability of the annotation process, a second expert annotator was employed to independently annotate a subset of the data. Specifically, four out of the 20 reviews previously annotated by the primary expert were selected, encompassing a total of 307 subtopics. The annotation process for both annotators was identical, ensuring consistency and allowing for a direct comparison of their results.
Agreement was assessed on three key tasks:

\subsubsection{Agreement for experiments in section 5.2}

\paragraph{Can LLMs effectively filter irrelevant subtopic clusters?}
The agreement between annotators regarding the relevancy of a subtopic to the review topic (title) achieved a 96\% percent agreement, where agreement is defined as the percentage of times raters agree. Annotators were asked to mark (1 or 0) if a subtopic is highly relevant and explicitly discuss all the terms in the topic.

\paragraph{Can LLMs organize subtopics into coherent thematic categories?} Thematic organization evaluation yielded another high percentage agreement of 96.5\%. The annotators were asked to mark (1 or 0) if a subtopic fit the assigned theme or not. A “fit” is defined if the theme correctly describes the membership of that subtopic. For example, a theme about brain pathologies would not fit a subtopic about brain control on the motor system, even if some of the papers in it deal with pathologies, while a subtopic about brain cancer will fit. Subtopics should be fully dedicated to the theme they belong to.

\subsubsection{Agreement for experiments in section 5.3}

End-to-End Subtopic Coverage Evaluation For subtopic coverage, we took the set of covered headers in the original table of contents in each review and measured the Jaccard similarity, meaning that we measured the number of agreed covered headers divided by the union of all annotated covered headers. We got an average of 95.4\% overlap coverage. This means that, on average, annotators agreed on most of the covered headers in the reviews.

These results imply that, unlike subjective annotation, this type of annotation is based on a common understanding of scientific terms and therefore achieves high agreement.

\newpage
\section{Prompts and system implementation details}
\subsection{Cluster Reader prompt}
\label{app:prompt_cluster_reader}
\begin{figure}[H]
\begin{boxA}
\ttfamily
\# Task Overview: \\
You are provided with a general topic and a set of scientific papers retrieved by a lexical search system using this topic as a query. Your task is to analyze how the papers relate to the topic and categorize their relevance. \\

\# Instructions: \\

Evaluate Relevance: Determine if the papers are directly related to the research topic. \\

- If they are not related to the research domain or do not address the topic directly, mark them as "NOT RELATED." \\
- If they are a genuine subtopic of the main topic, mark them as "RELATED." \\
- If the papers would not be relevant to a user searching for the main topic, consider them not related. \\
- IF the papers do not address an explicit relation to the topic, consider them not related. \\

\# Output Requirements: \\
Output should be a json with the following fields: \\
Description: Write a summary describing the common subtopic reflected in the research theme of the papers in the group in relation to the Topic. \\
Subtopic: Give a title for the group of papers that represents a meaningful subtopic of the Topic. \\
Relatedness: Rate the relatedness on a scale from 1 to 5, where 1 means not relevant at all, and 5 indicates the papers deal directly with the topic. \\
Is Related: State whether the papers are "RELATED" or "NOT RELATED" based on their relevance to the original topic. \\
- Write nothing else\\

Topic: \{query\}\\
Papers: \{papers\_list\}
\end{boxA}

\end{figure}
\FloatBarrier
\newpage
\subsection{Thematic Organization}
\label{app:prompt_thematic_org}
\begin{figure}[H]
\begin{boxA}
\ttfamily
You are given a nested dictionary where each key is a subtopic\_id and the value is a dictionary of subtopics of the topic \{query\}. Reflect on the subtopics and their descriptions and define clusters of topics that group the subtopics into meaningful research clusters. Create the clusters as an outline where each cluster is a foundational chapter about \{query\}. Those clusters will be used by a user to navigate between different domains of his research topic. Give each topic a clear label and describe the subtopics that the cluster is dealing with. Output must be in json. Do not leave any subtopic without a cluster. \\
\\
\#\# Output \\
- Output a json object with: \\
- clusters: list of dictionaries with digits from '1' to 'N' of "cluster\_ids", "cluster\_title" and "description" \\
- subtopics: dictionary with the subtopic\_id as a field and the appropriate cluster id as a key for each subtopic in the input.\\

Subtopic dictionary: \{subtopic\_and\_cluster\_ids\}
\text{  } 
\end{boxA}
\end{figure}
\FloatBarrier

\subsection{Subtopic Expander}
\label{app:prompt_subtopic_expander}
\begin{figure}[H]
\begin{boxA}
\ttfamily
You are tasked with identifying keywords or terms from a list of scientific paper titles and abstracts specifically relevant to the subtopic \{subtopic\_title\}. These keywords will be used to generate a query to retrieve documents about this specific subtopic. It's crucial to focus only on this particular aspect of \{query\} research and not on the general topic.

Here is the list of titles and abstracts:

Query Description: \{subtopic\_description\}

<titles\_and\_abstracts>\{papers\}</titles\_and\_abstracts>

Please follow these steps to complete the task:
1. Carefully read through each title and abstract.

2. As you read, identify words or short phrases that are specifically related to {subtopic}.

3. Pay special attention to recurring terms or concepts across multiple papers, as these are likely to be particularly relevant.

4. Avoid selecting overly general terms related to \{query\}. Focus on terms that specifically relate to the \{subtopic\_title\} aspect.

5. After analyzing all the titles and abstracts, compile a list of the most relevant and frequently occurring keywords or phrases.

6. Present your final list of keywords in order of relevance, with the most important or frequently occurring terms first. Include this list within <keywords> tags.

7. Provide a brief explanation for why you chose each keyword, highlighting its relevance to \{subtopic\_title\}. Include this explanation within <justification> tags.

Remember, the goal is to identify terms that will help retrieve documents specifically about \{subtopic\_title\}, not about \{query\} in general. Your selected keywords should reflect this focused approach.
"
\text{  } 
\end{boxA}
\end{figure}
Output:
\begin{figure}[H]
\begin{boxA}
\ttfamily
Subtopic Query : [Subtopic title + term$_{1}$,+...+,term$_{n}$]
\text{  } 
\end{boxA}
\end{figure}
\FloatBarrier

\twocolumn
\onecolumn
\section{\scitoc{} Examples and error analysis}
\subsection{\scitoc{} example}
\label{scitoc_example}
\begin{figure}[h]
    \centering
    \begin{tabular}{ll}
        &\textbf{Title: Visual Dysfunction in Diabetes} \\\\
        \textbf{2.} & \textbf{VISUAL DYSFUNCTION IN EARLY DIABETES} \\
        \textbf{3.} & \textbf{RETINAL NEURONAL DYSFUNCTION AND DEATH IN EARLY DIABETES} \\
        \hspace{0.2cm}\textbf{3.1.} & Changes in the Retinal Electroretinogram in Diabetes \\
        \hspace{0.2cm}\textbf{3.2.} & Changes in Retinal Neuronal Structure in Early Diabetes \\
        \textbf{4.} & \textbf{MECHANISMS OF RETINAL NEURONAL DYSFUNCTION IN EARLY DIABETES} \\
        \hspace{0.2cm}\textbf{4.1.} & Changes in Retinal Neuronal Inhibition in Early Diabetes \\
        \hspace{0.2cm}\textbf{4.2.} & Changes in Retinal Neuronal Glutamate Signaling in Early Diabetes \\
        \hspace{0.2cm}\textbf{4.3.} & Changes in Retinal Dopaminergic Signaling in Early Diabetes \\
        \hspace{0.2cm}\textbf{4.4.} & Diabetes May Have Distinct Effects on Retinal Pathways \\
        \hspace{0.2cm}\textbf{4.5.} & Potential Treatments Related to Neuronal Dysfunction \\
    \end{tabular}
    \caption{An example of the table of contents from the review ‘Visual Dysfunction in Diabetes’. The header for the ‘1. Introduction’ section has been removed from the benchmark.}
\end{figure}

\subsection{Error analysis} To better understand cases where Knowledge Navigator did not match TOC headers, we conducted an error analysis on the 20 annotated reviews, categorizing a total of 58 errors into two groups:

    \noindent\textbf{\textit{Grouping strategy mismatch} (50\% of errors )}: These cases represent discrepancies in how subtopic boundaries are defined. The Knowledge Navigator might include a subtopic within a broader category, while the review paper presents it as a standalone header. This highlights nuanced differences in how our system and human reviewers conceptualize and organize information.

    \noindent\textbf{\textit{Misses} (50\% of errors)}: These are relevant subtopics expected to appear in the system's output but were not found. This can occur for several reasons: relevant papers might be dispersed across multiple clusters, diminishing their individual impact; the initial corpus might lack sufficient coverage of the subtopic; or, in one isolated instance, a potentially matching subtopic was incorrectly filtered out by the subtopic filtering process.

\subsection{Header types removed from evaluation}
\label{app:non_valid_headers}
\begin{table}[h]
    \centering
    \small
    \begin{tabular}{| p{4cm} | p{5cm} | p{6cm} |}
        \hline
        \textbf{Header Type} & \textbf{Query} & \textbf{Header Example} \\ \toprule
        Introduction type headers & Tissue Immunity in the Bladder & 2.1. Anatomy \\ \hline
        Introduction type headers & Bacteriophages in the Dairy Industry & 3. THE PHAGE PROBLEM \\ \hline
        Subjective header & Schwann Cells & 4.1. Pathology Due to Defects in Canonical Functions \\ \hline
        Subjective header & APOBEC3 in Human Papillomavirus Infection and Oncogenesis & 4. CRITICAL GAPS IN OUR UNDERSTANDING OF APOBEC3 IN VIRUS-INDUCED CANCERS \\ \hline
    \end{tabular}
    \caption{Non Valid headers from \scitoc{} removed from the evaluation}
    \label{tab:headers_queries}
\end{table}

\newpage
\newpage
\onecolumn
\begin{table}[h]
\centering
\caption{Example of Knowledge Navigator output for the review Endocrine Disorders and COVID-19}
\begin{tabular}{|p{6cm}|p{10cm}|}
\toprule
\textbf{Themes} & \textbf{Subtopic Title and Description} \\
\midrule
\textbf{Impact of COVID-19 on General Endocrine Health} & \textbf{Impact of COVID-19 on Metabolic and Endocrine Health} \newline 
The common subtopic in these papers revolves around the intersection of COVID-19 and endocrine/metabolic disorders. The papers specifically address how metabolic syndrome, diabetes, and other related endocrine dysfunctions affect susceptibility to COVID-19, the severity of the disease, and the clinical management of these patients during the pandemic. \\
\cmidrule{2-2}
 & \textbf{Impact of COVID-19 on Adrenal Insufficiency and Glucocorticoid-related Endocrine Disorders} \newline 
The papers collectively discuss the intersection of endocrine disorders, particularly adrenal insufficiency and glucocorticoid-related diseases, with COVID-19. They examine the outcomes, management strategies, risk factors, and potential new onset of endocrine disorders in COVID-19 patients. \\
\midrule
\textbf{Thyroid Disorders and COVID-19} & \textbf{Thyroid Disorders Post COVID-19 Vaccination} \newline 
The papers predominantly discuss various thyroid disorders such as thyroiditis, thyrotoxicosis, and Graves' disease occurring after COVID-19 vaccination. They present case reports, studies, and reviews examining the potential link between COVID-19 vaccines and the onset or exacerbation of these endocrine conditions. \\
\cmidrule{2-2}
 & \textbf{Thyroid Dysfunction in COVID-19} \newline 
The common subtopic in these papers is the prevalence, impact, and outcomes associated with thyroid dysfunctions in patients who have contracted COVID-19. They explore various dimensions including the changes in thyroid hormone levels, the association of thyroid disorders with COVID-19 severity and outcomes, and potential mechanisms linking thyroid function with the disease. \\
\bottomrule
\end{tabular}
\end{table}

\section{Subtopic Expander experiments}
\subsection{Expert Relevancy Annotation and LLM-Judge}
\label{app:llm_judge}
We randomly sampled 13 subtopics out of the pool, and for each, we sampled 10 papers out of 100 to represent papers from the entire rank distribution, ending with 130 scientific papers. For each paper, the expert judged a score from 0 to 2, where "0" means the paper's relevancy to the subtopic is marginal, and "2" means the paper is focused on the subtopic. The annotation instructions were identical to those given to the LLM (see \ref{app:prompt_subtopic_expander}). We then let an LLM judge each paper in the same manner and evaluate their agreement. Since in some cases the degree of relevancy can be subjective between "2" and "1," we binarized the scores for "relevant" [2,1] and "not relevant" [0], similar to other recent studies on LLM relevancy judgment \citep{faggioli2023perspectives,thomas2024large}. We found that the binary agreement between the expert and the LLM reaches 87\% with a Cohen's kappa of 0.63, which is on par with other strong LLM relevancy judgment methods on TREC benchmarks \citep{thomas2024large}.
\end{document}